\documentclass[aps,twocolumn,usenatbib,nofootinbib,showkeys,showpacs,altaffilletter]{revtex4}

\usepackage{graphicx}
\usepackage{dcolumn}
\usepackage{bm}
\usepackage{latexsym}
\usepackage{mathrsfs}

\begin{document}


\title{The Closed String Tachyon and its relationship with the evolution of the Universe}

\author{Celia Escamilla-Rivera}
\email{celia_escamilla@ehu.es}
\affiliation{Fisika Teorikoaren eta Zientziaren Historia Saila, Zientzia
eta Teknologia Fakultatea, Euskal Herriko Unibertsitatea, 644 Posta
Kutxatila, 48080, Bilbao, Spain.}

\author{In collaboration with: G. Garcia Jimenez$^{1}$, O. Loaiza-Brito$^{2}$ and O. Obregon$^{2}$}
\affiliation{$^{2}$Facultad de Ciencias F\'isico Matem\'aticas, 
Universidad Aut\'onoma de Puebla, P.O. Box 1364, 72000, Puebla, Mexico.\\
$^{3}$Departamento de F\'isica, Divisi\'on de Ciencias e Ingenier\'ia, Campus Le\'on, Universidad de
Guanajuato, P.O. Box E-143, 37150, Le\'on, Guanajuato., Mexico.}
 

\begin{abstract}

We present a cosmological landscape where the classical closed string tachyon field plays an important
role in the framework of a critical bosonic compactification. Our cosmological solutions for a universe with
constant curvature describes an finite inflationary stage which expands till a maximum value before undergoes a
big crunch as the tachyon reaches the minimum of its potential. 

\end{abstract}


\maketitle


\section{Introduction}

The tachyon field cosmological models inspired by string theory has generated a great deal of
interest  over the last few years (see for example Refs.~\cite{Bergman:2006pd}). Some progress has been made in 
developing a effective theory in the context of open strings, taking 
into account the minimal state of energy towards the tachyon rolls down to the minimum of the potential, point 
in where the perturbative approach of the theory becomes reliable. This process is so-called tachyon 
condensation. Nonetheless, it has been proposed that tachyon condensation could provide a sinewy 
resolution of cosmological singularities \cite{Adams:2001sv}.

However,  we reminding the reader that tachyonic modes are not exclusively related to open strings only. 
Closed string tachyon has also generated a great speculation. This kind of modes arise in the bosonic 
string spectrum and at orbifold singularities within the context of superstring theory. The process  
which leads the closed string tachyon condensation is nicely described by a tachyonic 
potential which form has been inferred from String Field Theory (SFT)
and it is given by a nicely potential form $V(T)\sim -c_1^2T^2+c_2^2T^4$, where the maximum of the 
potential occurs at $T=0$. At this point we can ask whether any effects can arise when the source of 
tachyonic modes disappear, i.e., if the space-time can collapse itself. It is then natural to imagine an scenario 
with this kind of potential and expected that can be related with an inflationary universe which collapses as 
the tachyon condensates. Studies in this matter have been performed through the last few years as 
was showed in Refs.~\cite{Yang:2005rw}. Nonetheless, in this models the expansion stage is absent. 
Therefore, our goal was to found the scenario where an accelerated expansion can be possible.


\section{Effective closed string tachyon cosmology}\label{closestring}

Our alternative starting point is considering the low energy limit of critical bosonic string 
theory (which implies keeping the dilaton constant $\Phi_{0}$ and is not dynamically coupled to the
scale factor). In this situation we compactified on a 22-dimensional manifold. All this would imply the 
following Lagrangian\cite{EscamillaRivera:2011di}  
\begin{eqnarray}
\mathcal{L}&=&\frac{{m_p}^2}{2}\sqrt{-g^{E}} [R^{E}_{4} +6(\nabla\ln{\Omega})^2 +6\nabla^2 \ln{\Omega}
+4(\nabla\Phi)^2   \nonumber \\ &&
 -(\nabla T)^2-2\Omega^2 (\Psi,\psi)\mathcal{V}(T)],
\end{eqnarray}\label{general action}
where $T$ is the closed string tachyon field and $\mathcal{V}(T)$ is the effective tachyonic 
potential $\mathcal{V}$ as $\mathcal{V}(T)= V(T) -\frac{1}{2}R$. $g^{E}$ is 
the metric in Einstein frame and $\Omega$ is a conformal transformation from string to Einstein frame. 
Further analysis about it can be found in Ref.~\cite{EscamillaRivera:2011di}.
In the rest of this brief paper let us consider the tachyon field to be a function only on time, i.e. $T=T(t)$. Also  
consider the usual spatially flat FLRW metric $ds_{E}^{2}=-dt^{2}+e^{2\alpha(t)}\left(dr^{2}+r^{2}d{\Omega}^{2}\right)$. 
With these arrangements we compute 
the corresponding equations of motion in the Hamilton-Jacobi formalism,
\begin{equation}\label{eq:hamiltonian}
-2\left(\frac{\partial \mathcal{H}}{\partial T}\right)^2 +3\mathcal{H}^2 =\mathcal{V}(T).
\end{equation} 
As a next step we want to study under which conditions this formulation describes an universe which 
inflates and collapses as the tachyon runs down to the minimum of its potential, as expected from the
closed string tachyon framework. 


\section{Inflation stage}\label{GEvo}

Recall that the Hubble parameter $H$ determines the factor scale dynamics and choosing a specific
tachyon potential constraints it. Without losing this thought, we consider a polynomial $H$ in order to
have a SFT-potential of the form: $H(T)=-\frac{1}{2}(A+BT^2)$. The main point in this line is when $A=0$,
in where the SFT-potential is compatible with a effective scalar field only if the internal manifold is flat. In
this case the universe acquire a scale factor $a(t)=exp\left(-\frac{\Omega}{8} e^{4\Omega Bt}\right)$, where for $B>0$ 
describes an universe which starts at $t \rightarrow -\infty$ with $a=1$ and it always contracts. 

Now, within this context, we are capable to compute in the general case the necessary and sufficient 
conditions for inflation given by   $\epsilon\ll1$ and  $\eta\ll1$ where
\begin{eqnarray}
 \epsilon(T)
 &=&2\left(\frac{H'}{H}\right)^2=\frac{8 T^2}{(A/B +T^2)^2 }, \\
  |\eta(T)|&=&\left|\frac{1}{{\mathcal{V}}}\frac{\partial^2 {\mathcal{V}}}{\partial T^2}\right| =\frac{|(3A-4B)B+9B^2T^2|}{|\frac{3}{4}(A+BT^2)^2-2B^2T^2|}. \quad
 \label{epsilon}
 \end{eqnarray}
Two important aspects to be mentioned form here is that:
\begin{itemize}
\item From string theory it is expected that $A>>1$, i.e., high curvature of the internal space. This follows from the 
fact that ${R}\sim A^2\sim R^{-6}$, where $R$ is  the size of the extra dimensions.
\item Fixing $A$ and $B$ for guaranteed an inflation stage implies that the ratio between this values determines
how fast the universe expands/collapses.
\end{itemize}


\section{Collapse stage}

According to the standard cosmological model, assuming an universe filled by a perfect fluid EoS $p=w\rho$
in our closed tachyon landscape can show that $p$ is maximum at the minimum of the $\mathcal{V}(T)$. The
closed tachyon EoS is 
\begin{equation}
w=\frac{\frac{1}{2}\dot{T}^2 -\mathcal{V}(T)}{\frac{1}{2}\dot{T}^2 +\mathcal{V}(T)}=
-\frac{8B^2 (\Omega^2 +1)T^2 -3(A+BT^2)}{8B^2 (\Omega^2 -1)T^2 +3(A+BT^2)}.
\end{equation}
Interestingly, for early time, $w=-1$ and the tachyon potential evolves as a cosmological constant.

Regarding some specific conditions, we observed that for a particular form of $\mathcal{V}(T)$ the 
inflationary/expansion/collapse scenario is possible. More details are exposed in Ref.~\cite{EscamillaRivera:2011di}.


\section*{Acknowledgments}

C. E-R thanks to R. Jantzen, K. Rosquist and R. Ruffini for their invitation and support in the MG13 meeting 
and Paulo Moniz chairman in the Session CM4: Quantum Cosmology and Quantum Effects in the Early 
Universe for his advices and invitation. C. E-R is supported by Pablo Garc\'ia Fundation, FUNDEC, Mexico and the 
Department of Theoretical Physics UPV/EHU Research Group 317207ELBE.



\end{document}